%% file: main.tex
\documentclass[preprint,12pt]{elsarticle}

\usepackage{amsmath,amssymb,amsfonts,booktabs,hyperref,siunitx}

\usepackage{xurl}
\newcommand{\code}[1]{\nolinkurl{#1}}

\makeatletter
\def\ps@pprintTitle{%
  \let\@oddhead\@empty
  \let\@evenhead\@empty
  \def\@oddfoot{\centerline{\thepage}}%
  \let\@evenfoot\@oddfoot}
\makeatother 

\journal{Computer Physics Communications}

\begin{document}

\begin{frontmatter}



\title{G4HepEm: A Geant4 Extension for Efficient Electromagnetic Shower Simulation in High-Energy Physics}

\author[affil:cern]{Mih\'aly Nov\'ak\corref{cor1}}
\ead{mihaly.novak@cern.ch}
\cortext[cor1]{Corresponding author.}
\affiliation[affil:cern]{organization={European Organization for Nuclear Research (CERN)},
            addressline={Esplanade des Particules~1},
            city={1217~Meyrin},
            state={Geneva},
            country={Switzerland}
           }

\author[affil:cern]{Jonas Hahnfeld}
\ead{jonas.hahnfeld@cern.ch}

\author[affil:uwarwick]{Ben Morgan}
\ead{ben.morgan@warwick.ac.uk}
\affiliation[affil:uwarwick]{organization={University of Warwick, Department of Physics},
            addressline={Coventry CV4 7AL},
            country={UK}
           }

\begin{abstract}
Tracking electrons, positrons and photons dominates the CPU cost of detailed Geant4 detector simulation for the Large Hadron Collider (LHC) experiments. G4HepEm specialises this part of a Geant4 simulation with dedicated tracking loops and a compact, stateless electromagnetic physics library that can run on both CPU and GPU, while geometry and navigation remain native Geant4. Sensitive detectors and user actions are invoked through the standard Geant4 interfaces, allowing application code that relies on these interfaces to be used unchanged. Optional algorithmic performance optimisations, such as Woodcock tracking of photons and a combined multiple-scattering and transportation algorithm, provide further speedup. G4HepEm reproduces native Geant4 results in a sampling calorimeter and the CMS High-Granularity Calorimeter test beam while simulating electron showers at speedups of 2.1–2.2$\times$. In full-detector simulations of top-quark pair events, it reduces the simulation time by 14.1\% (CMS Run~3), 21.0\% (preliminary CMS Run~4) and 16.3\% (ATLAS Run~3), while simulated CMS hit distributions and reconstructed ATLAS observables remain in close agreement with native Geant4. G4HepEm is open source and integrated into the CMS and ATLAS software frameworks, where it can contribute significantly to the experiments' efforts to cope with the computing challenges of the High-Luminosity LHC (HL--LHC) era.
 \end{abstract}




\end{frontmatter}



\section{Introduction}
\label{sec:introduction}
ATLAS \cite{ATLAS2008} and CMS \cite{CMS2008} are the largest High Energy Physics (HEP) experiments at the Large Hadron Collider (LHC), designed independently to record collision data using their general-purpose detectors, which cover a wide range of physics phenomena. Beyond the recorded collision events, a large number of simulated events are produced, which are essential for interpreting and analysing the experimental data.

This Monte Carlo (MC) production begins with \textit{event generation}, where primary particles from simulated collisions are created. This is followed by \textit{detector simulation}, which models the interactions of the particles with the detector materials and produces energy deposits in sensitive detector components. Next comes  \textit{digitisation}, which simulates the detector electronics' response to these energy deposits, producing simulated  raw data signals. These are then processed by the same \textit{reconstruction} software used for real detector outputs (see \cite{ATLAS_simulation_2010,boudoul2015monte} for more details).

Among the individual steps of this MC production, detector simulation is typically one of the most computationally  demanding. Prior to Run 3, approximately 38\% of the total CPU hours consumed by ATLAS computing activities were devoted to detector simulation---more than event generation, reconstruction, and analysis combined. Even under these conditions, the sensitivity of some analyses remained limited by the available statistics of simulated events \cite{ATLAS_HLLHC_computing_2020}.

This challenge is expected to intensify in the upcoming High-Luminosity LHC (HL--LHC) era (Runs 4 and 5), which aims to deliver approximately 3000~fb$^{-1}$ of integrated luminosity---roughly 10 times more than the combined 300--500 fb$^{-1}$ of Runs~1--3 \cite{hilumi-overview,arxiv2505.03535}. Since the number of simulated events typically exceeds the recorded ones by a factor of 2--3 (to reduce statistical uncertainties), the HL--LHC era not only substantially extends the physics potential but also introduces unprecedented computing challenges.

According to a recent report \cite{ATLAS_SoftComp_Roadmap_2022}, ATLAS must undertake aggressive computing R\&D during the preparation for Run 4 to address this challenge. Improving the performance of their detector simulation is a crucial element of this programme, given its dominant contribution to the overall computing workload. In contrast, detector simulation has not been the primary computational bottleneck in CMS, in part due to a less demanding detector configuration. 
However, this situation is expected to change following the Run 4 detector upgrade. The introduction of the High-Granularity Calorimeter (HGCAL) \cite{French_HGCAL_2025} is projected to increase CMS detector simulation time by a factor of three \cite{CMS_Computing_Phase2_2022} compared to Run 3. CMS plans to mitigate this slowdown through a twofold performance improvement targeted for the start of Run 4, achieved through substantial dedicated R\&D on its detector 
simulation, similarly to ATLAS.

Producing the large number of simulated events required for physics analyses during the HL--LHC era solely with traditional, detailed simulation is no longer feasible due to its prohibitive computing cost. Consequently, the LHC experiments plan to meet their Run 4 computing targets by combining full simulation with a range of fast simulation techniques. Replacing the accurate but computationally demanding step-by-step modelling of radiation--matter interactions with faster, approximate, yet sufficiently accurate methods under specific conditions is not new to the HL--LHC\footnote{For example, ATLAS used the fast \textit{Frozen Shower} library \cite{barberio2008fast} to simulate low-energy electromagnetic showers in the forward calorimeter region even before Run 3 \cite{ATLAS_HLLHC_computing_2020}.}. However, from Run 4 onward, the experiments will be required to employ and rely on such techniques to a much greater extent due to these computational constraints \cite{ATLAS_SoftComp_Roadmap_2022,CMS_Computing_Phase2_2022}. Fast simulation must generally be accurate enough that the same correction factors and uncertainties can be applied as for detailed simulation \cite{ATLAS_HLLHC_computing_2020}. To this end, the LHC experiments follow two complementary strategies: (i) optimise the computing performance of their detailed detector simulation to maximise its share of the simulated samples, and (ii) improve the accuracy of their fast simulation models so they can provide the remaining required events.

Geant4 \cite{agostinelli_geant4simulation_2003,allison_geant4_2006,allison_recent_2016} is the de facto simulation toolkit employed by all HL--LHC experiments for detailed detector modelling. G4HepEm is a Geant4 extension that provides alternative electron, positron, and gamma transport simulation specialised for HEP detector simulations. It offers substantial performance improvement in full detector simulation while maintaining the modelling accuracy and detail. Consequently, G4HepEm can play a pivotal role in supporting the HL--LHC experiments’ efforts to optimise detailed simulations and address the computing performance challenges anticipated for Run 4.

\section{Motivation}
\label{sec:motivation}
Geant4 \cite{agostinelli_geant4simulation_2003,allison_geant4_2006,allison_recent_2016}  is a generic simulation toolkit for modelling the passage of radiation through matter. Despite the wide range of particle types, radiation forms, and physical interaction characteristics it supports, the toolkit employs a single generic tracking algorithm. The algorithm is sufficient to calculate all the details required to perform a simulation step---independent of particle type or interaction---while covering a broad variety of interaction phenomena with significantly different characteristics. This generality is achieved by relying on carefully designed interfaces, managers and intermediate objects during the step computation.

However, in the case of LHC detector simulations, the fraction of computing time spent on tracking just three particle types---electrons, positrons, and gammas---can reach 60--80 \%, depending on the experiment and simulation configuration. Moreover, the corresponding physics required for modelling is not only well defined, but also forms a small closed set for electromagnetic shower generation, and is common across all LHC experiments.
Therefore, replacing the generic tracking loop with specialised algorithms---one for electrons and positrons, and another for gammas---coupled with an efficient implementation of the common physics required by HEP experiments, offers the potential for substantial improvements in detector simulation performance without compromising accuracy. Part of the gain comes from removing unused flexibility in the generic design---avoiding interfaces, managers and intermediate objects during the step computation---and instead providing an efficient streamlined code dedicated to these particles. The dedicated tracking loops also enable the integration of optional, particle-type-specific algorithmic optimisations that can deliver substantial performance benefits under specific conditions. 
These include Woodcock tracking for gamma photons and the combined multiple Coulomb scattering (MSC) with transportation for electrons and 
positrons, both described in the next section.

\section{Implementation}
\label{sec:implementation}

\begin{figure}[tbp]
\centering
\includegraphics[width=\columnwidth]{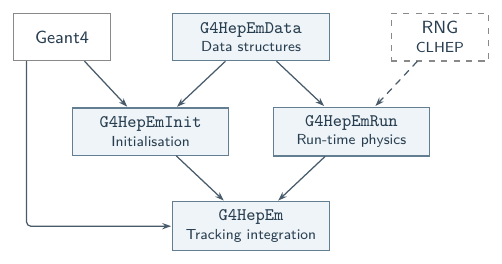}
\caption{Organisation of the four G4HepEm libraries and their principal dependencies. Arrows point from a dependency to the library that uses it. The dashed arrow denotes the replaceable random-number generator, supplied by CLHEP in Geant4 applications.}
\label{fig:architecture}
\end{figure}

A key design principle of G4HepEm is the explicit separation of initialisation and run-time functionalities. Initialisation requires access to a large and diverse set of data and algorithms that are not needed once event processing begins. Since the physics data structures are populated at initialisation and only read at run time, this naturally leads to separating their definitions from the functional code. As a result, G4HepEm is organised into four libraries: \code{G4HepEmData}, \code{G4HepEmInit}, \code{G4HepEmRun} and \code{G4HepEm}. Their organisation and principal dependencies are illustrated in Fig.~\ref{fig:architecture}.

\code{G4HepEmData} contains only the definitions of the data structure---including cross-section, stopping-power, range, and other physics tables---that are required for the physics modelling at run time. Data structures use a compact, cache-efficient memory layout that reflects their runtime usage patterns and avoids unnecessary indirections. 

These data can then be initialised using the functionalities implemented in \code{G4HepEmInit}. All necessary data are extracted from the appropriate Geant4 EM physics models using the corresponding interfaces, making the initialisation dependent not only on the data definitions themselves but also on Geant4. 

Depending only on the data definitions, \code{G4HepEmRun} then implements the complete set of EM physics functionalities required for and actually utilised by the LHC detector simulations. Functionalities offered by this lightweight run-time library, together with the physics data extracted at initialisation, were selected to be equivalent to those available in the Geant4 standard EM physics constructor. The corresponding physics models and energy ranges, together with the differences from the Geant4 standard EM configuration, are summarised in Table~\ref{tab:physics-models} in~\ref{app:physics-models}. Additional options, such as the relaxed multiple-scattering step-limit algorithm used by CMS and LHCb, are also included, extending the G4HepEm capabilities towards Geant4 EM standard physics options 1 and 2. These alternative algorithms, similarly to other multiple scattering parameters that further control the related step limit, can be selected per detector region which allows G4HepEm to reproduce the fine-tuned EM physics configurations utilised by the different LHC experiments. The available configuration settings and their global or region-dependent scope are summarised in Table~\ref{tab:configuration} in~\ref{app:physics-models}.

The \code{G4HepEm} library integrates the physics functionality provided by \code{G4HepEmRun} with Geant4's native geometry and navigation to form complete, dedicated stepping and tracking loops for electrons/positrons and photons. These loops are implemented in \code{G4HepEmTrackingManager}, which implements the \code{G4VTrackingManager} interface introduced in Geant4 version~11.0 to support the requirements of G4HepEm. This interface allows the generic tracking and stepping machinery to be replaced for selected particle types. G4HepEm can therefore be integrated into a Geant4 application by constructing a \code{G4HepEmTrackingManager} and attaching it to the \code{G4Electron}, \code{G4Positron} and \code{G4Gamma} particle definitions through \code{G4ParticleDefinition::SetTrackingManager}.

For these particles, Geant4 hands each \code{G4Track} directly to the attached tracking manager, which assumes full responsibility for its propagation and interactions. The native \code{G4TrackingManager} and \code{G4SteppingManager} are bypassed while the native Geant4 user interfaces are invoked during the specialised tracking/stepping. Therefore, the application code that retrieves the current track or step through these native \code{G4TrackingManager} and \code{G4SteppingManager} must be adapted. Furthermore, the native Geant4 processes registered to the particle are omitted as their functionalities are replaced by those provided in G4HepEm. Photo-nuclear and electron-/positron-nuclear interactions receive special treatment: G4HepEm stores their interaction cross sections and includes them in the interaction sampling, but delegates final-state generation to the corresponding native Geant4 process when one of these relatively rare interactions occurs.

Beyond integration with Geant4, the separation of initialisation and run-time functionality enables standalone use of the physics library. The initialised physics data and parameters can be saved to a JSON file and restored in subsequent runs, avoiding the Geant4-dependent initialisation procedure. Together with \code{G4HepEmRun}, this restored state provides the electromagnetic physics functionality needed to build a standalone shower simulator. This capability has been utilised in \code{HepEmShow} \cite{hepemshow}, a standalone C++ electromagnetic shower simulation for a simplified sampling calorimeter, used to study the differentiability of electromagnetic showers \cite{aehle2025optimization,aehle2024efficient}.

The same separation also supports GPU execution. Core physics functions in \code{G4HepEmRun} are implemented in headers and operate on basic particle quantities and physics data, without depending on a particular track representation. When built with CUDA support, the initialised data structures can be transferred to device memory, allowing the corresponding run-time functions to operate on these data directly on the GPU. This provides the same electromagnetic physics implementation on both host and device, while leaving track management and the execution model to the transport framework. The AdePT project uses this capability to employ G4HepEm as its electromagnetic physics engine for GPU-based detector simulation \cite{amadio2023offloading}.

\subsection{Special tracking options for performance optimisation}
\label{sec:special_tracking_options}
Having the special electron/positron and gamma tracking loops implemented locally in \code{G4HepEmTrackingManager} not only streamlines the related particle tracking but also offers the possibilities to implement special, optional tracking algorithms that can provide significant additional performance benefits in some special cases. G4HepEm offers such options both for gamma and electron/positron tracking.

\subsubsection{Woodcock or delta tracking of photons}
\label{sec:special_tracking_options_Woodcock}
Photon interaction probabilities are determined by macroscopic cross sections that depend on the photon energy and tracking material. Since the energy remains constant between interactions, these cross sections change only at material boundaries. Because materials are assigned to geometrical volumes in Geant4, conventional tracking ends simulation steps at volume boundaries and updates the remaining interaction length using the properties of the material entered. In highly granular geometries, photon mean free paths can greatly exceed individual volume dimensions, resulting in numerous boundary crossings before a physical interaction.

Woodcock, or delta, tracking avoids these boundary-limited steps by introducing a material-independent majorant that bounds the total macroscopic cross sections throughout a selected detector region \cite{woodcock1965techniques, coleman1968mathematical}. The difference between the majorant and the local real physical cross section defines a fictitious interaction cross section. At a given photon energy, the real and fictitious cross sections therefore sum to the same majorant in every material of the selected region. Candidate interaction distances can thus be sampled without stopping at internal volume boundaries. At each candidate point, a physical interaction is accepted with probability equal to the local real total cross section divided by the majorant; otherwise, the photon continues with unchanged energy and direction. This technique is particularly beneficial in granular geometries where the navigation savings outweigh the cost of fictitious interactions.

As part of the work presented here, one of the authors provided an early implementation of Woodcock tracking in the Geant4 version used by ATLAS during Run~3. This allowed the experiment to benefit from this technique ahead of the availability of G4HepEm already during the Run~3 production, yielding a substantial performance improvement of approximately 18\% when applied to their electromagnetic endcap calorimeter (EMEC)~\cite{atlas_run3_computing_2025}. However, that implementation required departures from Geant4's usual tracking conventions, particularly the responsibility of transportation for geometrical propagation, and is restricted to a single detector region. The dedicated photon tracking loop in \code{G4HepEmTrackingManager}, which also handles transportation, enables a more natural and general implementation, supporting Woodcock tracking independently in multiple detector regions without requiring these departures.

\subsubsection{Combined multiple Coulomb scattering (MSC) and transportation for electrons and positrons}
\label{sec:special_tracking_options_Combined}

For electrons and positrons, the cumulative effect of many Coulomb scatterings over a transport step is described by an MSC model. For accurate modelling of the resulting angular deflection and lateral displacement, MSC may impose a step length shorter than the distance to either the next volume boundary or the next discrete interaction. Several successive steps may therefore be limited by MSC; these are processed as separate Geant4 steps in the native tracking. Combining MSC and transportation can therefore reduce the overhead of handling the two processes separately. Although this is already true for a single step, the main computational benefit is obtained when several successive steps are limited by MSC as these steps can then be performed internally, without returning control to the tracking loop or exposing each intermediate endpoint to the user interfaces. At each internal substep, the particle is transported and its kinetic energy is updated while the angular deflection and lateral displacement due to MSC are applied. The transport continues from the updated position and direction, while the true path length and mean energy loss are accumulated over the substeps. The sequence ends when another limit is reached, such as a volume boundary or a discrete interaction, or when the particle stops. A single Geant4 step is then reported to the application, with the final position, direction and kinetic energy of the particle, the accumulated total path length and the energy deposited over the entire step.

When the readout segmentation does not coincide with the tracking geometry, the energy deposited during a step may need to be distributed among several readout cells. If the path is approximated by the straight line joining the pre- and post-step points, the spatial distribution of the deposit is approximate even for a single native Geant4 step in the presence of MSC. Combining several MSC-limited substeps can amplify this effect: their intermediate positions are used during transport but are not reported to the scoring code. The straight line between the endpoints of the combined step may therefore cross different cells, or traverse them in different proportions, than the sequence of substeps. The multiple-substep option should therefore be disabled in detector regions where scoring relies on this straight-line approximation unless its effect on the relevant observables has been shown to be negligible. The ATLAS calorimeter results in Section~\ref{sec:results_ATLAS} illustrate this sensitivity.

As part of this work, combined MSC and transportation was made available in Geant4~11.1 through the \code{G4TransportationWithMsc} process~\cite{hahnfeld2024geant4}. In G4HepEm, the algorithm is implemented directly in the electron and positron tracking loop of \code{G4HepEmTrackingManager}. Multiple internal substeps are enabled by default, but can be disabled independently for each detector region. In that case, MSC and transportation remain combined and each MSC-limited step is reported separately.

\section{Results}
\label{sec:results}

The physics modelling and computational performance of G4HepEm are evaluated using applications of increasing complexity. Two standalone Geant4 applications are considered first: a simplified sampling calorimeter, which provides a controlled setting for physics validation, and the CMS HGCAL test-beam setup~\cite{acar2023performance,cms2022response}. As part of this work, G4HepEm was integrated into Athena~\cite{atlas_athena} and CMSSW~\cite{cms_cmssw}, the software frameworks of ATLAS and CMS, respectively. These integrations allowed G4HepEm to be evaluated using the complete detector geometries of both experiments and the detector-simulation settings used in Run~3 production. A preliminary Run~4 configuration is also studied for CMS.\footnote{A corresponding ATLAS Run~4 configuration was not available at the time of this study.}

All timing measurements were performed on the same system, equipped with two AMD EPYC~7302 processors (32 physical cores and 64 hardware threads in total) and running AlmaLinux~9. For both standalone applications, static Release builds with GCC~13.1.0 and Geant4~11.2.2 were used, and the simulations were run with 64 worker threads. Within each benchmark, native Geant4 and G4HepEm were compared under the same execution conditions.

\subsection{Simplified sampling calorimeter}
\label{sec:results_Simplified_sampling_calorimeter}

The Geant4 \code{TestEm3} example, a configurable sampling calorimeter regularly used by Geant4 developers to test electromagnetic physics for HEP applications, is used to establish a baseline for the physics and performance comparisons. Its simple geometry, consisting of alternating layers of absorber and active material, allows the comparison to focus on the physics modelling before more complex detector geometries are considered. The calorimeter response and execution time obtained with G4HepEm are compared with those of the corresponding native Geant4 simulation.

A modified version of \code{TestEm3} is provided among the G4HepEm examples, with G4HepEm integrated through a local EM physics constructor as described in Section~\ref{sec:implementation}. This constructor is combined with \code{G4EmExtraPhysics} in a modular Geant4 physics list to include photonuclear and electron-/positron-nuclear interactions. The calorimeter is configured with 50 layers, each consisting of 2.3~mm of $\mathrm{PbWO}_4$ and 5.7~mm of liquid argon, and is irradiated with a monoenergetic 10~GeV electron beam. The \code{G4HepEmTrackingManager} is configured to use Woodcock tracking for photons within the calorimeter and the default multiple-step configuration of combined MSC and transportation for electrons and positrons. A 0.7~mm secondary production cut is used, with the threshold applied to all EM interactions.\footnote{setting \texttt{/process/em/applyCuts true}, as in the LHC simulations.}  

\begin{figure}[tbp]
\centering
\includegraphics[width=\columnwidth]{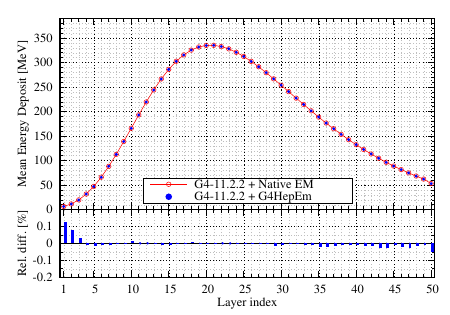}
\caption{Longitudinal energy-deposition profiles for 10~GeV electrons in a simplified sampling calorimeter consisting of 50 layers, each comprising 2.3~mm of $\mathrm{PbWO}_4$ and 5.7~mm of liquid argon. Results obtained with G4HepEm are compared with those from native Geant4, with $2\times10^7$ primary electrons simulated in each case.}
\label{fig:simplified_sampling_calorimeter_Profile}
\end{figure}

For the reference simulation, the Geant4 EM physics constructor is configured to match the G4HepEm model selection, using the Urban MSC model also above 100~MeV (see Table~\ref{tab:physics-models}). With this configuration, excellent agreement is observed between the longitudinal energy-deposition profiles obtained with G4HepEm and native Geant4, as shown in Fig.~\ref{fig:simplified_sampling_calorimeter_Profile}. Using the unmodified Geant4 standard EM constructor, which employs the WentzelVI MSC model above 100~MeV, introduces somewhat larger relative differences, reaching $0.05$--$0.1\%$. These small differences reflect the alternative MSC model configurations, both of which are suitable for the application considered.

\begin{table}[t]
\centering
\caption{Shower observables and recorded step counts in the simplified sampling calorimeter for $2\times10^7$ primary electrons of 10~GeV. All quantities are means per primary. 
}
\label{tab:simplified_sampling_calorimeter_Performance}
\begin{tabular}{lrrr}
\hline
Quantity & Geant4 & G4HepEm & $\Delta$ [\%] \\
\hline
\multicolumn{4}{l}{Mean energy deposit [MeV]} \\
\quad $\mathrm{PbWO}_4$ & 6703.98 & 6703.78 & $+0.003$ \\
\quad Liquid argon      & 2535.71 & 2535.95 & $-0.009$ \\
\hline
\multicolumn{4}{l}{Mean number of secondaries} \\
\quad $\gamma$ & 4344.88 & 4346.72 & $-0.042$ \\
\quad $e^-$    & 1735.35 & 1733.80 & $+0.089$ \\
\quad $e^+$    &  426.06 &  425.99 & $+0.016$ \\
\hline
\multicolumn{4}{l}{Mean number of recorded steps} \\
\quad Charged & 26933.5 & 12150.4 & $+54.89$ \\
\quad Neutral & 40228.5 &  8413.8 & $+79.08$ \\
\hline
\end{tabular}
\end{table}

The mean energy deposited in each material and the mean numbers of secondary photons, electrons and positrons agree within $0.1\%$ as reported in Table~\ref{tab:simplified_sampling_calorimeter_Performance}. In contrast, the mean numbers of recorded neutral and charged steps are reduced by approximately $80\%$ and $55\%$, respectively. The reduction in neutral steps results from Woodcock tracking, which eliminates simulation steps imposed by volume boundaries within the selected calorimeter region. The reduction in charged steps is the result of enabling the combined MSC and transportation for multiple MSC-limited steps: MSC-limited substeps are performed internally rather than being exposed as separate tracking steps. These reductions are therefore the result of the optional tracking optimisations, while close agreement is maintained for all physics-related observables.

The timing comparison is performed separately using $10^6$ primary electrons of 10~GeV. Disabling Woodcock tracking and the multiple-substep option of the combined MSC and transportation, \code{G4HepEmTrackingManager} reduces the run time by approximately $25\%$ through the specialised tracking and the G4HepEm physics implementation. A dedicated benchmark using specialised tracking loops that directly invoke the native Geant4 physics processes gives a reduction of approximately $12\%$, indicating the contribution from specialised tracking alone. Enabling Woodcock tracking within the calorimeter then increases the reduction from approximately $25\%$ to $41\%$,  while additionally enabling multiple internal MSC-limited substeps within the combined MSC and transportation increases it further to approximately $51\%$. In this fully optimised configuration, the elapsed run time decreases from 2192.8~s for native Geant4 to 1065.9~s for G4HepEm, corresponding to a speedup of approximately $2.1$.

\subsection{CMS HGCAL test-beam application}
\label{sec:results_HGCALTB}

As discussed in Section~\ref{sec:introduction}, the CMS High-Granularity Calorimeter (HGCAL) upgrade is expected to introduce substantial additional computational demands for detailed detector simulation during the HL--LHC era \cite{CMS_Computing_Phase2_2022}. The HGCAL test-beam setup used at CERN in 2018 \cite{acar2023performance,cms2022response} consists of an electromagnetic section (CE-E) with 28 silicon sampling layers interleaved with lead, copper and copper-tungsten absorbers, a hadronic section (CE-H) with 12 silicon sampling layers and steel absorbers, and a downstream CALICE analogue hadronic calorimeter (AHCAL). The fine segmentation, detailed module structures and numerous material boundaries make this geometry considerably more complex than the simplified sampling calorimeter, providing an intermediate benchmark before considering the complete CMS or ATLAS detectors.

The standalone Geant4 implementation of this test-beam setup, developed for Geant-val, was adapted to integrate G4HepEm and used in this study\footnote{\url{https://github.com/geant-val/HGCALTB}}. The \code{G4HepEmTrackingManager} is configured with Woodcock tracking enabled for photons within the calorimeters, while the default multiple-substep configuration of combined MSC and transportation is used for electrons and positrons, as in the simplified sampling calorimeter.

\begin{figure}[tbp]
\centering
\includegraphics[width=\columnwidth]{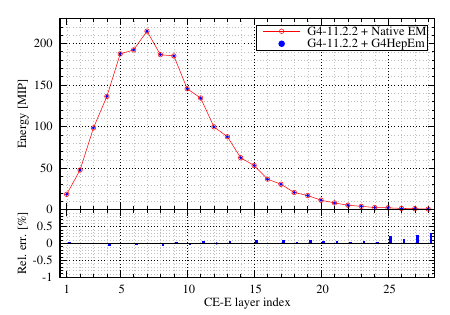}
\caption{Longitudinal energy-deposition profiles for 20~GeV electrons in the CMS HGCAL test-beam setup. Results obtained with G4HepEm are compared with those from native Geant4, with $10^7$ primary electrons simulated in each case.}
\label{fig:hgcal_Profile}
\end{figure}

The longitudinal energy-deposition profiles obtained with G4HepEm and native Geant4 for 20~GeV primary electrons are compared in Fig.~\ref{fig:hgcal_Profile}. Excellent agreement can be observed throughout the electromagnetic calorimeter with relative differences within approximately $0.1\%$ over most of the shower length with somewhat larger differences appearing towards the end of the calorimeter, where the deposited energy is small.

The execution times are compared separately using $3.2\times10^4$ primary particles of 20~GeV for both electrons and negatively charged pions. For electrons, the elapsed run time decreases from 304.87~s for native Geant4 to 136.75~s for G4HepEm, corresponding to a run-time reduction of approximately $55\%$, or a speedup of $2.2$. For pions, the run time decreases from 118.54~s to 74.24~s, corresponding to a reduction of approximately $37\%$, or a speedup of $1.6$.

The larger performance improvement for electrons reflects the electromagnetic nature of their showers, whereas pion showers also include a hadronic component. Since G4HepEm accelerates only the transport of electromagnetic particles, the overall run-time reduction is smaller for pions.

\subsection{LHC experiments: CMS}
\label{sec:results_CMS}

G4HepEm was already available as an external library in CMSSW~\cite{cms_cmssw}. As part of this work, the library was updated to tag \code{20250220}, and a dedicated electromagnetic physics constructor was implemented based on the native \code{CMSEmStandardPhysics} constructor. The new constructor attaches \code{G4HepEmTrackingManager} to the electron, positron and photon definitions, leaving the transport of all other particles unchanged. The integration was incorporated into the CMSSW~15.1 development series. The results presented here were obtained with \code{CMSSW\_15\_1\_0\_pre1}, using \code{FTFP\_BERT\_EMM} as the native Geant4 reference and \code{FTFP\_BERT\_EMH} for the G4HepEm simulation.

The G4HepEm physics settings were chosen to reproduce those of the native CMS electromagnetic constructor, including the regional treatment of multiple Coulomb scattering. The relaxed \code{fMinimal} step limitation was used throughout most of the detector, while the more conservative \code{fUseSafety} algorithm was used in the hadron calorimeter (HCAL). In the preliminary Run~4 configuration, the conservative MSC step limitation was also applied in HGCAL. Photon Woodcock tracking was enabled in G4HepEm only in HGCAL and hence only for Run~4.

The comparison was performed with the \code{CaloSimHitStudy} application using $10^5$ $t\bar{t}$ events for each detector and transport configuration. For Run~3, the 2024 detector geometry based on DD4hep was used with proton--proton collisions at $\sqrt{s}=13.6~\mathrm{TeV}$. For Run~4, the preliminary D110 geometry, including HGCAL, was used with collisions at $\sqrt{s}=14~\mathrm{TeV}$. The physics comparison was made at the level of simulated hits.

\begin{figure}[tbp]
\centering
\includegraphics[width=\columnwidth]{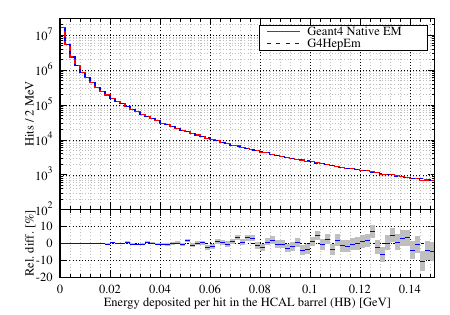}
\par\smallskip
\includegraphics[width=\columnwidth]{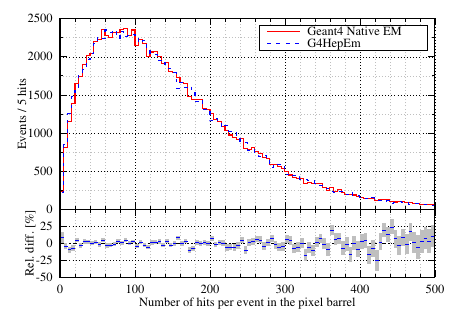}
\caption{Energy deposited per hit in the HCAL barrel (top) and number of hits per event in the pixel barrel (bottom) for $10^5$ $t\bar{t}$ events at $\sqrt{s}=14~\mathrm{TeV}$ in the preliminary CMS Run~4 geometry. The relative differences are shown as $100(N_{\mathrm{ref}}-N_{\mathrm{G4HepEm}})/N_{\mathrm{ref}}$, where $N_{\mathrm{ref}}$ is the native Geant4 bin count. Grey bars indicate the statistical uncertainties on the differences.}
\label{fig:cms_run4_hits}
\end{figure}

Comparable agreement with native Geant4 was observed for the Run~3 and Run~4 configurations. Two representative Run~4 distributions are shown in Fig.~\ref{fig:cms_run4_hits}: the energy deposited per hit in the HCAL barrel and the number of hits per event in the pixel barrel. The HCAL barrel spectrum is reproduced over several orders of magnitude in hit count, while the pixel-barrel distribution shows close agreement through both the peak and the falling part of the multiplicity spectrum. In both observables, the bin-to-bin fluctuations increase towards the less populated tails, as do the statistical uncertainties. No clear systematic shift is observed in either distribution.

\begin{figure}[tbp]
\centering
\includegraphics[width=\columnwidth]{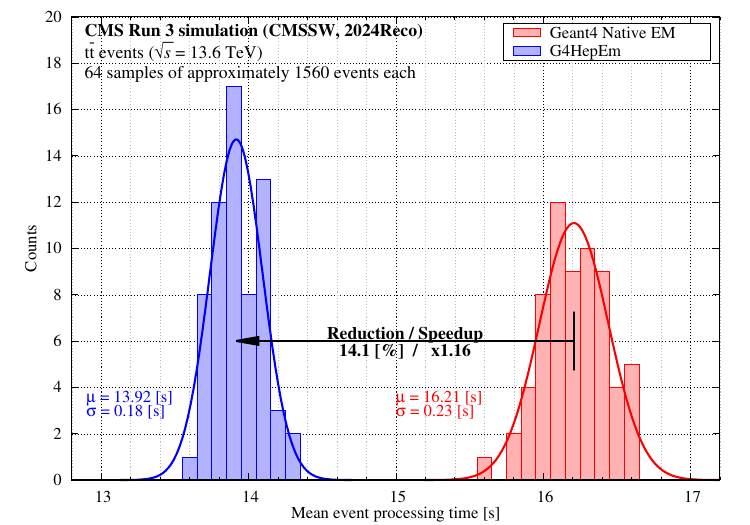}
\par\smallskip
\includegraphics[width=\columnwidth]{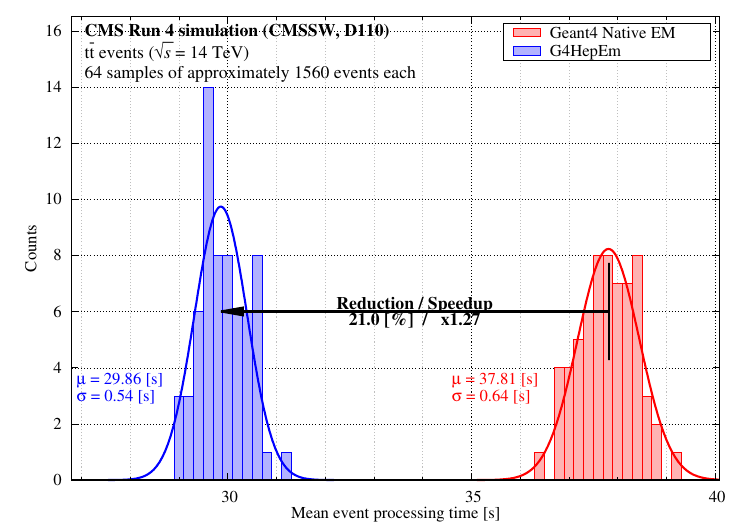}
\caption{Mean detector-simulation time per $t\bar{t}$ event in CMSSW for the Run~3 configuration at $\sqrt{s}=13.6~\mathrm{TeV}$ (top) and the preliminary Run~4 configuration at $\sqrt{s}=14~\mathrm{TeV}$ (bottom). Each histogram entry is the mean reported by one worker thread. The curves are Gaussian fits, and the run-time reduction and speedup are calculated relative to the native Geant4 reference in each configuration.}
\label{fig:cms_timing}
\end{figure}

Detector-simulation times were measured using 64 worker threads for each sample of $10^5$ events. Each worker reported its mean event processing time over approximately 1560 events. The distributions of these 64 thread means and the corresponding Gaussian fits are shown in Fig.~\ref{fig:cms_timing}. For Run~3, the fitted mean decreased from 16.21~s with native Geant4 to 13.92~s with G4HepEm, corresponding to a run-time reduction of $14.1\%$, or a speedup of $1.16$. For Run~4, the fitted mean decreased from 37.81~s to 29.86~s, corresponding to a reduction of $21.0\%$, or a speedup of $1.27$.

The Run~3 gain is substantial given that the crystal electromagnetic calorimeter and the relaxed MSC step limitation permit relatively long electron and positron steps. A larger gain was expected for Run~4 because of the finer HGCAL segmentation, the more conservative MSC step limitation and the opportunity to use Woodcock tracking. Although the absolute saving increased from 2.29~s to 7.95~s per event, the relative reduction increased by only 6.9 percentage points. Profiling provided an explanation: approximately 3.5 times as many tracks were produced in the Run~4 configuration as in Run~3, with a large contribution from low-energy secondary electrons in HGCAL. In the D110 setup studied here, a uniform secondary-production range cut of $30~\mu\mathrm{m}$ was applied throughout HGCAL.

The effect of this cut was examined separately in the standalone HGCAL test-beam application using 20~GeV primary electrons and the same $30~\mu\mathrm{m}$ range cut. Of the approximately $1.69\times10^4$ secondary electrons produced in HGCAL per primary, $93\%$ never left the volume in which they were created, and $69\%$ completed their transport in a single step. Such tracks incur the cost of track creation and stacking while offering little opportunity to benefit from faster stepping. This helps explain why the relative performance gain did not increase in proportion to the Run~4 tracking workload. It also identifies the uniform HGCAL production cut as a candidate for optimisation: cuts tailored to the dimensions of different detector regions could reduce the number of short-lived tracks, subject to validation of the resulting detector response.

\subsection{LHC experiments: ATLAS}
\label{sec:results_ATLAS}

The complete detector geometry and simulation settings used in ATLAS Run~3 production were taken as the reference for this study~\cite{atlas_run3_computing_2025}. G4HepEm was added to a local build of the ATLAS external software stack based on \code{atlasexternals} version~2.1.16
and integrated into Athena version~24.0.42%
. 
The same ATLAS-patched Geant4~11.2.2 build was used for the reference and G4HepEm simulations. The ATLAS-specific transition-radiation process was retained for the transition radiation tracker (TRT), together with support for native fast-simulation processes and Athena's sensitive-detector and user-action interfaces.

As discussed in Section~\ref{sec:special_tracking_options_Woodcock}, Woodcock tracking of photons in the electromagnetic endcap calorimeter (EMEC) was already used in Run~3 production and was therefore enabled in the native Geant4 reference. In the test configuration, native Geant4 transport was replaced by G4HepEm for electrons, positrons and photons by attaching \code{G4HepEmTrackingManager} to their particle definitions through a dedicated EM physics constructor in the \code{FTFP\_BERT\_ATL\_HepEm} physics list. All other particles were transported as in the reference. For photons, G4HepEm Woodcock tracking was enabled in both the EMEC and the electromagnetic barrel calorimeter (EMB). For electrons and positrons, combined MSC and transportation was used with its default multiple-substep option enabled.

For physics validation, $9\times10^4$ single-lepton $t\bar{t}$ events from proton--proton collisions at $\sqrt{s}=13.6~\mathrm{TeV}$ were simulated without pile-up.
The settings followed those used by ATLAS for its validation of Geant4~11.2.2. For each transport configuration, the events were processed through detector simulation, digitisation and reconstruction using the ATLAS physics-validation workflow~\cite{ATLAS_simulation_2010}, allowing the effects of the transport implementation to be assessed at the level of reconstructed observables.

\begin{figure}[tbp]
\centering
\includegraphics[width=\columnwidth]{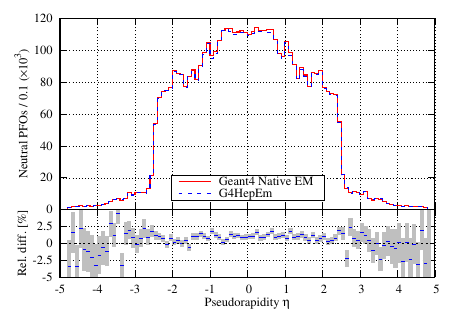}
\par\smallskip
\includegraphics[width=\columnwidth]{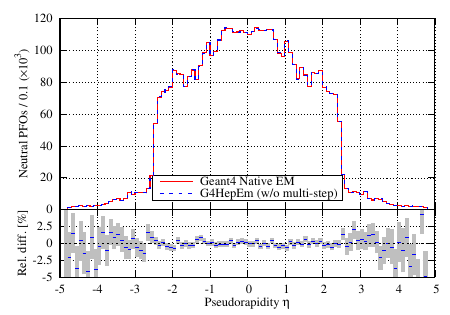}
\caption{Pseudorapidity distribution of reconstructed neutral particle-flow objects in ATLAS for $9\times10^4$ single-lepton $t\bar{t}$ events at $\sqrt{s}=13.6~\mathrm{TeV}$ without pile-up. Native Geant4 electromagnetic transport is compared with G4HepEm using its default multiple-substep mode (top) and with this option disabled (bottom).}
\label{fig:atlas_eta}
\end{figure}

\begin{figure}[tbp]
\centering
\includegraphics[width=\columnwidth]{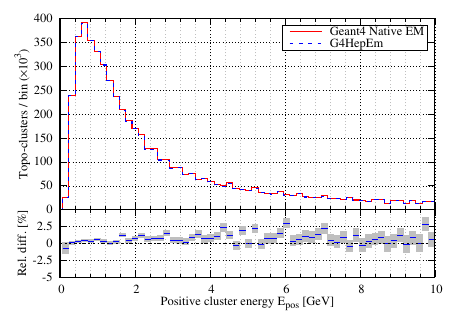}
\par\smallskip
\includegraphics[width=\columnwidth]{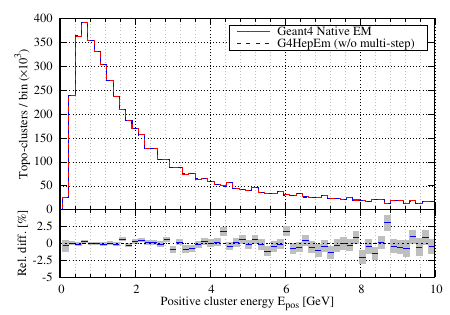}
\caption{Positive energy moment $E_{\mathrm{pos}}$ of calorimeter topological clusters in ATLAS for the same event sample as in Fig.~\ref{fig:atlas_eta}. Native Geant4 electromagnetic transport is compared with G4HepEm using its default multiple-substep mode (top) and with this option disabled (bottom).}
\label{fig:atlas_engpos}
\end{figure}

Close agreement with the native Geant4 simulation was observed in general, with minor differences limited only to a few sensitive observables. Two examples are shown in Figs.~\ref{fig:atlas_eta} and~\ref{fig:atlas_engpos}: the pseudorapidity distribution of neutral particle-flow objects~\cite{ATLASParticleFlow2017} and the positive energy moment of calorimeter topological clusters~\cite{ATLASTopoClusters2017}. The former probes the reconstructed neutral-object yield across the detector acceptance; the latter, $E_{\mathrm{pos}}$, measures the positive cell-energy contributions assigned to each cluster. The relative-difference panels show $100(N_{\mathrm{ref}}-N_{\mathrm{G4HepEm}})/N_{\mathrm{ref}}$ for each bin. The grey bars show the $\pm1\sigma$ statistical uncertainty on the relative difference, propagated from the reference and test histograms assuming independent samples.

In the ATLAS calorimeters affected, the readout segmentation in $\eta$ and $\phi$ does not coincide with the tracking geometry. The sensitive-detector code therefore distributes the energy deposit reported by Geant4 in a given simulation step among the cells between the pre- and post-step points assuming a straight-line path. However, the actual path is not only unknown when using an MSC model for describing multiple Coulomb scattering, but can be very different from the straight line joining the pre- and post-step points. When multiple MSC-limited steps are combined and only the cumulative step is exposed to the user code, with the corresponding first pre- and last post-step points, the straight-line path approximation can become even less adequate, leading to the observed differences. Disabling the multiple-substep option restored close agreement with the native Geant4 simulation for all observables, consistent with this interpretation. 

\begin{figure}[tbp]
\centering
\includegraphics[width=\columnwidth]{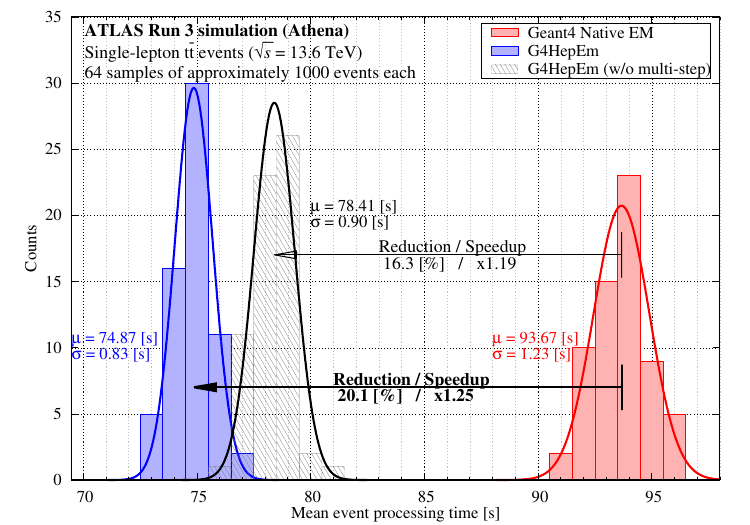}
\caption{Mean detector-simulation time per event for the ATLAS Run~3 configuration. Each entry is the mean reported by one worker thread for approximately 1000 single-lepton $t\bar{t}$ events at $\sqrt{s}=13.6~\mathrm{TeV}$. The native Geant4 reference (red) includes Woodcock tracking in the EMEC. G4HepEm is shown with its default multiple-substep mode enabled (blue) and disabled (hatched). The curves are Gaussian fits.}
\label{fig:time_ATLAS_Run3}
\end{figure}

Detector-simulation performance was measured separately using $64\,000$ single-lepton $t\bar{t}$ events per configuration. Eight Athena processes were run concurrently, each with eight worker threads and an independent sample of 8000 events. Within each process, the events were distributed among the workers, so each thread reported a mean processing time based on approximately 1000 events. The resulting 64 thread means per configuration are shown in Fig.~\ref{fig:time_ATLAS_Run3}. The widths of the fitted Gaussian distributions were approximately $1.1$--$1.3\%$ of their respective means, indicating little variation among the thread averages.

Relative to the native Geant4 reference with EMEC Woodcock tracking already enabled, the mean simulation time per event decreased from 93.67~s to 74.87~s with G4HepEm in its default multiple-substep mode. This corresponds to a further run-time reduction of $20.1\%$, or a speedup of $1.25$. With multiple internal substeps disabled, the mean time was 78.41~s: the configuration that restored close agreement for the sensitive calorimeter observables still provided a $16.3\%$ reduction relative to the same reference, or a speedup of $1.19$.

\section{Conclusions}
\label{sec:conclusions}

G4HepEm is an open-source library for the specialised transport of electrons, positrons and photons in Geant4-based detector simulation. Its dedicated tracking loops combine compact, stateless electromagnetic physics functions with native Geant4 geometry and navigation. Sensitive detectors and other user code are invoked through the standard Geant4 interfaces. Moreover, G4HepEm makes the same electromagnetic physics available on the GPU. The AdePT project utilises this capability for a GPU-based detector simulation, with \code{G4HepEmTrackingManager} serving as its CPU fallback~\cite{amadio2023offloading}.

In the simplified sampling calorimeter and the CMS HGCAL test-beam setup, excellent agreement with native Geant4 was observed in the longitudinal energy-deposition profiles, with relative differences of about 0.1\% or less over most of the shower. Electron-induced showers were simulated 2.1--2.2 times as fast. In the sampling calorimeter, enabling Woodcock tracking for photons and multiple internal MSC substeps for electrons and positrons increased the run-time reduction from about 25\% to 51\%, relative to the same native Geant4 reference. The dependence of the gain on the electromagnetic fraction of the workload was illustrated by the test-beam results: a speedup of 2.2 was obtained for electrons, compared with 1.6 for pions.

Integration into the CMS and ATLAS software frameworks allowed these gains to be assessed for complete collision events. For $t\bar{t}$ events in CMS, the mean detector-simulation time was reduced by 14.1\% for Run~3 and 21.0\% for a preliminary Run~4 configuration, with close agreement in the simulated hit distributions examined. Profiling of the Run~4 configuration identified a large population of short-lived, low-energy secondary electrons produced with the uniform $30~\mu\mathrm{m}$ production cut in HGCAL. This suggests that region-dependent cuts merit investigation, subject to validation of the detector response. For ATLAS Run~3, disabling multiple internal MSC substeps restored close agreement for reconstructed observables sensitive to calorimeter scoring, while retaining a 16.3\% reduction in simulation time.

For the same detector-simulation CPU budget, the measured speedups correspond to approximately 19.5\% more fully simulated $t\bar{t}$ events for the ATLAS Run~3 configuration and 26.6\% more for the preliminary CMS Run~4 configuration. By reducing the cost of detailed detector simulation, G4HepEm complements the experiments' fast-simulation programmes and helps address the computing demands of the HL--LHC.

\section*{Software availability}

The G4HepEm source code, including the \code{TestEm3} sampling-calorimeter example, is available under the Apache License~2.0 at \url{https://github.com/Geant4/g4hepem}. Installation instructions and documentation of the physics models and library interfaces are available at \url{https://g4hepem.readthedocs.io/}.

\section*{CRediT authorship contribution statement}

\textbf{Mihály Novák:} Writing -- original draft, Visualization, Software, Methodology, Investigation, Conceptualization.
\textbf{Jonas Hahnfeld:} Writing -- review \& editing, Software, Methodology, Investigation, Conceptualization.
\textbf{Ben Morgan:} Writing -- review \& editing, Software.

\section*{Declaration of competing interest}

The authors declare that they have no known competing financial interests or personal relationships that could have appeared to influence the work reported in this paper.

\section*{Data availability}

Data will be made available on request.

\section*{Acknowledgements}

The authors thank the ATLAS and CMS simulation teams for their support during the integration of G4HepEm into Athena and CMSSW, respectively, and for their help in verifying the resulting implementations.

\appendix
\section{Summary of G4HepEm physics and configuration options}
\label{app:physics-models}
\setcounter{table}{0}

Table~\ref{tab:physics-models} summarises the physics models used in G4HepEm and their relation to the Geant4 standard EM configuration while the available configuration settings and their global or region-dependent scope are summarised in Table~\ref{tab:configuration}.

\input{appendix_1_g4hepem_physics_table}

\input{appendix_1_g4hepem_configuration_table.tex}

\section*{Declaration of generative AI and AI-assisted technologies in the manuscript preparation process}

ChatGPT (OpenAI) was used to help revise the wording of parts of the manuscript. The authors reviewed and edited the text and take full responsibility for its content.

\bibliographystyle{elsarticle-num} 
\bibliography{main}

\end{document}

%% file: appendix_1_g4hepem_physics_table.tex

\begin{table*}[t]
\centering
\caption{Physics models used in G4HepEm compared with the Geant4 standard
EM configuration (option~0).
The listed energy intervals refer to model coverage,
not to production cuts. Differences in the treatment of scattering and
photoelectric absorption are made explicit.}
\label{tab:physics-models}
\begingroup
\small
\setlength{\tabcolsep}{4pt}
\renewcommand{\arraystretch}{1.15}
\begin{tabular}{@{}p{0.07\textwidth}p{0.17\textwidth}p{0.25\textwidth}p{0.28\textwidth}p{\dimexpr0.23\textwidth-8\tabcolsep\relax}@{}}
\toprule
Particle & Interaction & Geant4 (option~0) & G4HepEm & Energy interval \\
\midrule
$e^-$ & Ionisation & M\o ller & M\o ller & 1~keV--100~TeV \\
$e^+$ & Ionisation & Bhabha & Bhabha & 1~keV--100~TeV \\
\addlinespace
$e^\pm$ & Bremsstrahlung & Seltzer--Berger & Seltzer--Berger & 1~keV--1~GeV \\
 & & Relativistic model\textsuperscript{a} & Relativistic model\textsuperscript{a} & 1~GeV--100~TeV \\
\addlinespace
$e^\pm$ & Coulomb scattering & Urban & Urban & 1~keV--100~MeV \\
 & & Wentzel-VI (mixed) & Urban & 100~MeV--100~TeV \\
\addlinespace
$e^+$ & Annihilation & $e^+e^-\rightarrow 2\gamma$ & $e^+e^-\rightarrow 2\gamma$ & At rest--100~TeV \\
\midrule
$\gamma$ & Photoelectric effect & Livermore & Simplified treatment\textsuperscript{b} & 0--100~TeV\textsuperscript{b} \\
\addlinespace
 & Compton scattering & Klein--Nishina\textsuperscript{c} & Klein--Nishina\textsuperscript{c} & 100~eV--100~TeV \\
\addlinespace
 & Pair production & Bethe--Heitler\textsuperscript{d} & Bethe--Heitler\textsuperscript{d} & $2m_ec^2$--100~TeV \\
\addlinespace
 & Rayleigh scattering & Livermore & Not included & 100~keV--100~TeV\textsuperscript{e} \\
\bottomrule
\end{tabular}

\vspace{4pt}
\begin{minipage}{\textwidth}
\footnotesize
The $e^\pm$ energy-loss fluctuation implemented in G4HepEm corresponds to \texttt{G4UniversalFluctuation} in Geant4~11.0.p01.
\par\smallskip
\textsuperscript{a}Bethe--Heitler differential cross section with screening
and Coulomb corrections, the Wheeler--Lamb treatment of emission in the
field of atomic electrons, and Landau--Pomeranchuk--Migdal (LPM) suppression.
\par
\textsuperscript{b}G4HepEm uses a simplified implementation of
\texttt{G4PEEffectFluoModel}, retaining only the K-shell binding energy.
Photons below the highest binding energy are absorbed without producing
a secondary photoelectron.
\par
\textsuperscript{c}Electron binding is accounted for through a scattering-function
correction to the integrated cross section, but not in final-state sampling.
\par
\textsuperscript{d}Includes screening and Coulomb corrections, conversion
in the field of atomic electrons, and LPM suppression for
$E_\gamma>100$~GeV.
\par
\textsuperscript{e}The interval applies only to the Geant4 entry.
\end{minipage}
\endgroup
\end{table*}

%% file: appendix_1_g4hepem_configuration_table.tex

\begin{table*}[t]
\centering
\caption{Global physics parameters and configuration options in G4HepEm.
The first group is inherited from \texttt{G4EmParameters} during
initialisation, except for the photon tracking cut. The remaining groups
are configurable through \texttt{G4HepEmConfig}. Regional settings can be
assigned separately for each detector region or uniformly across all
existing regions. MSC denotes multiple Coulomb scattering.}
\label{tab:configuration}
\begingroup
\small
\setlength{\tabcolsep}{5pt}
\renewcommand{\arraystretch}{1.12}
\begin{tabular}{@{}p{0.29\textwidth}p{0.15\textwidth}p{\dimexpr0.56\textwidth-4\tabcolsep\relax}@{}}
\toprule
Setting & Scope & Purpose \\
\midrule
\multicolumn{3}{@{}l}{\textit{Global physics parameters}\textsuperscript{a}} \\
\addlinespace
Electron tracking cut & Global & Sets the electron tracking energy threshold. \\
Photon tracking cut\textsuperscript{b} & Global & Sets the photon tracking energy threshold. \\
Energy-loss table limits & Global & Set the minimum and maximum energies of the energy-loss tables. \\
Energy-loss table binning & Global & Sets the number of energy-loss table bins. \\
MSC positron correction\textsuperscript{c} & Global & Enables or disables the positron correction to the MSC angular-width parameter $\theta_0$. \\
MSC lateral displacement & Global & Enables or disables displacement in MSC. \\
Linear energy-loss limit & Global & Sets the limit associated with the linear energy-loss approximation. \\
\midrule
\multicolumn{3}{@{}l}{\textit{Regional configuration options}} \\
\addlinespace
Energy-loss step limitation & Per region & Sets the fractional-range and final-range parameters of the continuous energy-loss step-limit function. \\
Energy-loss fluctuations & Per region & Enables or disables fluctuations in continuous energy loss. \\
MSC range factor & Per region & Sets the range-factor parameter used in MSC step limitation. \\
MSC safety factor & Per region & Sets the safety-factor parameter used in MSC step limitation. \\
MSC step-limit algorithm & Per region & Selects the minimal step-limit algorithm or the safety-based alternative. \\
Application of production cuts & Per region & Enables or disables applying production cuts to processes beyond ionisation and bremsstrahlung. \\
\midrule
\multicolumn{3}{@{}l}{\textit{Optional tracking algorithms}} \\
\addlinespace
Woodcock tracking: activation\textsuperscript{d} & Per region & Selects the detector regions in which photon Woodcock tracking is enabled. \\
Woodcock tracking: energy limit & Global & Sets the photon energy below which Woodcock tracking is disabled; the constructor sets 200~keV. \\
Combined MSC and transportation & Per region & Enables or disables multiple transportation steps within the combined MSC and transportation treatment for $e^-/e^+$. \\
\bottomrule
\end{tabular}

\vspace{4pt}
\begin{minipage}{\textwidth}
\footnotesize
\textsuperscript{a}These parameters have no dedicated setters in
\texttt{G4HepEmConfig}. Except for the photon tracking cut, their values
are copied from \texttt{G4EmParameters} during initialisation; subsequent
changes in Geant4 are not automatically propagated. The global linear
energy-loss limit is copied uniformly to all regions. Table limits and
binning must be set before the corresponding physics tables are built.
\par
\textsuperscript{b}The photon tracking cut is initialised to zero. A
nonzero value must be assigned directly to
\texttt{G4HepEmParameters::fGammaTrackingCut}; it is not inherited from
\texttt{G4EmParameters}.
\par
\textsuperscript{c}The MSC positron correction is inherited from
\texttt{G4EmParameters} for Geant4 versions 11.1 and later; for earlier
versions it is initialised to \texttt{true}.
\par
\textsuperscript{d}The Woodcock region list must be set before tracking
initialisation, immediately after construction of the tracking manager.
For the other regional settings with dedicated setters, the configuration
interface permits changes before the run begins.
\end{minipage}
\endgroup
\end{table*}